Figure 1

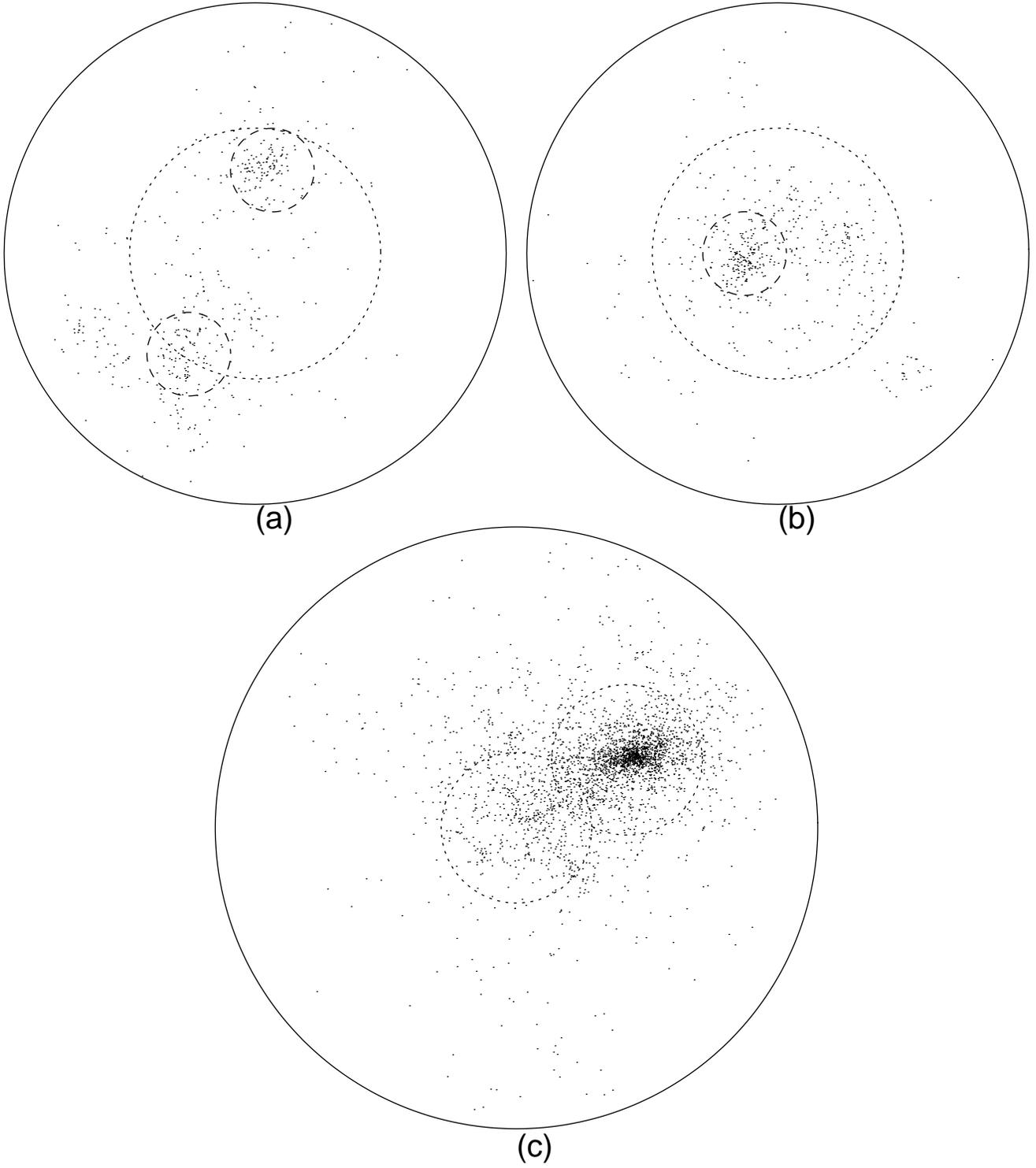

(a)

(b)

(c)

# Abundance of moderate-redshift clusters in the Cold + Hot dark matter model

Y.P. Jing[1] &  L.Z. Fang[1,2]

1 Department of Physics, University of Arizona, Tucson, 85721
2 Steward Observatory, University of Arizona, Tucson, 85721

## Abstract

Using a set of P$^3$M simulation which accurately treats the density evolution of two components of dark matter, we study the evolution of clusters in the Cold + Hot dark matter (CHDM) model. The mass function, the velocity dispersion function and the temperature function of clusters are calculated for four different epochs of $z \leq 0.5$. We also use the simulation data to test the Press-Schechter expression of the halo abundance as a function of the velocity dispersion $\sigma_v$. The model predictions are in good agreement with the observational data of local cluster abundances ($z = 0$). We also tentatively compare the model with the Gunn and his collaborators' observation of rich clusters at $z \approx 0.8$ and with the x-ray luminous clusters at $z \approx 0.5$ of the *Einstein* Extended Medium Sensitivity Survey. The important feature of the model is the rapid formation of clusters in the near past: the abundances of clusters of $\sigma_v \geq 700 \, \mathrm{kms}^{-1}$ and of $\sigma_v \geq 1200 \, \mathrm{kms}^{-1}$ at $z = 0.5$ are only 1/4 and 1/10 respectively of the present values ($z = 0$). Ongoing ROSAT and AXAF surveys of distant clusters will provide sensitive tests to the model. The abundance of clusters at $z \approx 0.5$ would also be a good discriminator between the CHDM model and a low-density flat CDM model both of which show very similar clustering properties at $z = 0$.

# 1. Introduction

After the COBE quadropole detection $Q = (6.0 \pm 1.5) \times 10^{-6}$ of the microwave background radiation (MBR) (Smoot et al. 1992), one popular model of the galaxy formation is the hybrid model (hereafter CHDM model) which assumes a flat ($\Omega = 1$) universe with dark matter composed of about 70% in Cold Dark Matter (CDM) and 30% in Hot Dark Matter (HDM) and assumes a Zeldovich spectrum $[P(k) \propto k]$ for the primordial density fluctuation. The model has been studied recently by many authors using analytical methods and N-body simulations, and the model predictions are compatible with observational data presently available (see Schaefer & Shafi 1993 for a review). Therefore the model is worthy of further investigation.

Cluster abundance constructs a useful test for models of galaxy formation. Using Particle-Mesh (PM) N-body simulation, Jing et al. (1993a, hereafter JMBF93) studied the mass function of rich clusters at $z = 0$ for this model. They found that the mass function is a bit higher than the observed one of Bahcall & Cen (1992, BC92), if the density spectrum is normalized to the COBE quadropole $Q_{COBE} = 6.0 \times 10^{-6}$. A similar conclusion has been drawn by Bartlett & Silk (1993) based on the Press-Schechter formula. They found that the model with the $Q_{COBE}$ normalization produces a temperature function of clusters (at $z = 0$) which is slightly higher than the observations of Edge et al. (1990) and Henry & Arnaud (1992). Both studies used the linear density power spectrum of Holtzman (1989) and suggested that a smaller normalization, e.g., $Q = 4.5 \times 10^{-6}$ which, however, is still within the $1\sigma$ error of the COBE quadropole, can better fit the cluster observations. As shown in §4, this reduction in the normalization may be unnecessary because the linear power spectrum $P(k)$ from a more accurate calculation of Klypin et al. (1993; hereafter KHPR93) is lower by $\sim 20\%$ on cluster mass scales than that of Holztman (1989) when the two power spectra are normalized equally on very large scale.

In this paper, we study the formation and evolution of clusters of galaxies in the CHMD model using a set of high resolution Particle-Particle/Particle-Mesh ($P^3M$) N-body simulations. Each simulation uses $64^3$ CDM particles and $2 \times 64^3$ HDM particles to follow density evolution in a cubic box of 128 $h^{-1}$Mpc on each side. From the simulation a set of clusters (and rich groups) are identified and the cluster properties are compared with available observations. This paper will concentrate on the abundance of clusters at the present as well as at moderate redshifts ($z \lesssim 0.5$).

As will be shown, the CHDM model is in good agreement with the abundances of local clusters (§4; see also Bartlett & Silk 1993). A low-density flat CDM model can equally well account for the observations (BC92, also §4). Indeed both models are very popular after the COBE $Q$ detection, and have been found rather successful in explaining many observational data of the large-scale structures of the universe (e.g., Davis et al. 1992; Jing et al. 1993a&b; KHPR93; Schaefer & Shafi 1993 and references therein; Wright et al. 1992; Efstathiou et al. 1992 and references therein). Therefore it is interesting to find tests to discriminate between these two models. The abundance of clusters at moderate redshift can serve such a test. Not only are the abundances quite different in the models, but also such observational data are being accumulated by some ongoing surveys, e.g., ROSAT and AXAF. In fact, there already exist observations on this subject. One is the faint cluster survey of Gunn et al. (GHO; 1986), another is the *Einstein* Extended Medium Sensitivity Survey (EMSS) of clusters (Henry et al. 1992).

In §2, we will give details about the cosmological model and the model parameters we used. Then we describe our techniques used in the N-body simulation. In order to compare the CHDM with other models, we have also run simulations for the Standard CDM model (SCDM) and for a low-density flat CDM model with a non-zero cosmological constant (LCDM). In §3, we will describe our method of identifying clusters from the simulations. Our identification procedure is divided into two steps. The first step is to find a cluster list with cluster members exceeding a certain number within a certain radius. The method is based on the *friends-of-friends* algorithm (e.g., White et al. 1987a; JMBF93). Although in most circumstances (clusters) the method works quite successfully, it will, as we will show, give misleading results in the case of complicated distribution of particles in the simulation. We remedy this by searching for gravitational potential minimums around the clusters (the second step). Then we will compare the model predictions in §4 with available observational data at $z = 0$, including the mass function, the cluster temperature functions and especially the new CfA data of cluster velocity dispersion function of Zabludoff et al. (1993; hereafter ZGHR93). While both the CHDM and the LCDM model can nicely fit the observations at $z = 0$, the CHDM shows a very rapid evolution of cluster abundance even within $z \leq 0.5$. Therefore it is very interesting to confront the model with observations at redshift $z \approx 0.5$. A comparison with six clusters at $0.40 < z < 0.60$ in the EMSS shows that the CHDM model predicts a right amount of high velocity dispersion clusters ($\sigma_v > 1500 \,\mathrm{kms^{-1}}$) at $z = 0.5$. However, if the abundance of rich clusters at $z \approx 0.8$ is as high as claimed by Gunn (1990), the CHDM model would be ruled out at a high confidence level. As an interesting by-product, our simulation provides a good database for testing the Press-Schechter (P-S) formula which has been frequently used to predict the abundance of clusters. Although there have been a number of simulation tests on the P-S prediction of abundance of collapsed objects $n(M)$ as a function of mass $M$, much less work has been done on the abundance $n(\sigma_v)$ as a function of velocity dispersion $\sigma_v$ (but see Narayan and White 1988, hereafter NW88). Since the relation between $M$ and $\sigma_v$ depends on cluster density profile, it's non-trivial to relate $M$ and $\sigma_v$. We will discuss this problem in §5. Finally in §6 we summarize our results.

## 2. Cosmological models and N-body simulations

We assume for the CHDM model that the universe is flat, has a zero cosmological constant $\Lambda_0 = 0$ and is composed in mass of 60% CDM ($\Omega_{\rm CDM} = 0.6$), 30% HDM ($\Omega_{\rm HDM} = 0.3$) and 10% baryon ($\Omega_{\rm baryon} = 0.1$). The present Hubble constant is taken to be $50 \, {\rm kms^{-1} \, Mpc^{-1}}$ or $h = 0.5$ (where $h$ is the Hubble constant in unit of $100 \, {\rm kms^{-1} \, Mpc^{-1}}$). Furthermore HDM is assumed to be made of one flavor of massive neutrinos (say $\tau$ neutrinos) and the other two flavors are taken to be massless. The mass of massive neutrino is $m_\nu = 6.9 \, {\rm eV}$. The neutrinos thermal motion satisfies the Fermi-Dirac distribution:

$$dn(v) \propto \frac{v^2 dv}{\exp[v/v_0(z)] + 1}, \qquad (1)$$

where $v_0(z) = (1+z)ckT_\nu/m_\nu$ and $T_\nu = 1.95K$. For this model a number of authors have calculated the transfer functions in the linear regime. We use the transfer functions of KHPR93 which gave the fitting formulae for CDM $[T_c(k,z)]$ and for HDM $[T_h(k,z)]$ at different cosmological redshift. Assuming the primordial fluctuation has a Zel'dovich spectrum $P(k) \propto k$, the processed linear spectra then are $P_c(k,z) \propto T_c^2(k,z)k$ for CDM and $P_h(k,z) \propto T_h^2(k,z)k$ for HDM. The spectra are normalized to give an *rms* quadropole of MBR $Q = Q_{COBE}$ as measured by the COBE. This corresponds to $\sigma_8 \approx 0.6$, where $\sigma_8$ is the present linearly evolved *rms* CDM density fluctuation in a sphere of radius $8 \, h^{-1} {\rm Mpc}$. Since the transfer functions for baryons and CDM are the same (KHPR93) and we don't treat baryon and CDM separately, we regard baryon and CDM as one component, and simply use subscript CDM for this component (e.g., $\Omega_{\rm CDM} = 0.7$ in the model now).

To facilitate comparison with other interesting models, we have also run simulations for the models of SCDM and LCDM. The SCDM model has been popular in the past decade (Davis & Efstathiou 1988), and the LCDM model has been considered as a successful model to account for most observational data of cosmological clustering (e.g., Wright et al. 1992; Efstathiou et al. 1992). The SCDM model is assumed to have $h = 0.5$, $\Omega_0 = \Omega_{\rm CDM} = 1$ and $\Lambda_0 = 0$; and the LCDM model has $h = 0.75$, $\Omega_0 = \Omega_{\rm CDM} = 0.3$ and $\Lambda_0 = 0.7$. For both models, we use the transfer functions of Bardeen et al. (1986), and also assume that the primordial power spectrum is the Zel'dovich one. The SCDM is normalized to have $Q = 3 \times 10^{-6}$, so that $\sigma_8 \approx 0.6$, close to (but a little higher than) the common normalization of the SCDM (Davis et al. 1985) and also close to the $\sigma_8$ of the CHDM model. The SCDM model is not motivated by the COBE $Q$ experiment, because it is well known the model normalized by $Q_{COBE}$ produces too many rich clusters. We normalize the LCDM model by $Q = 4.5 \times 10^{-6}$. This $Q$ value is smaller than $Q_{COBE}$ but equals to the $1\sigma$ lower limit of the COBE $Q$ measurement, so the model is still motivated by the COBE $Q$ experiment. Why we take the normalization of $Q = 4.5 \times 10^{-6}$ is that the model has $\sigma_8 = 1$ and that the model correctly produces the present abundance of rich clusters (see §4). If we used $Q = Q_{COBE}$ for the normalization, we would have to assume an anti-biasing distribution for galaxies at $r \sim 8\,h^{-1}$Mpc and the model would produce too many rich clusters at the present time.

Our simulations are performed by a P$^3$M code in a cubic box of $128^3\,h^{-3}$Mpc$^3$ with periodic boundaries. $128^3$ meshes are uniformly spread in the box for generating initial conditions and for calculating mesh force. For the CHDM model, $64^3$ CDM particles and $2 \times 64^3$ HDM particles are used to follow the density evolution in the simulation. Each CDM particle has mass of $3.1 \times 10^{12} M_\odot$ and each HDM particle has mass of $0.67 \times 10^{12} M_\odot$. Rich clusters, like Abell ones, will be composed of about 100 CDM particles and 200 HDM particles. Our code is designed according to the standard method described by Hockney & Eastwood (1981) and Efstathiou et al. (1985). Briefly, in the code, the triangular-shaped clouds (TSC) scheme is used for mass assignment and mesh force interpolation. The gravitational force is split into two parts: the mesh force and short range interaction force. The short range force is softened on separation less than $\eta = 0.1\,h^{-1}$Mpc by assuming that each particle has a radius $\eta/2$ and a linear radial density profile. We applied the short range force for particles with separation less than $2.7\,h^{-1}$Mpc, which would result in force discontinuity about 2% at $2.7\,h^{-1}$Mpc (Efstathiou et al. 1985). The mesh force is obtained from potential on the grid with the four-point finite-difference approximation. The potential is obtained by solving the Poisson equation using the FFT technique with the Green function given by Hockney & Eastwood (1981; §8.3). The motion of particles in the comoving cosmological coordinate is integrated using the leapfrog technique. The integration variable is the cosmological scale factor $a$, and the step size is taken to be 0.02 $a_i$ (where $a_i$ is the scale factor at the beginning of simulation). The simulations are started at redshift $z_i = 8$, so we need 400 integration steps to evolve the simulations to the present time. Five realizations are generated for the CHDM model.

The initial conditions for the CHDM simulation are generated in the same way as Jing et al. (1993b). The method is based on the Zel'dovich approximation but with a modification to consider the existence of HDM. It requires that 1.) the initial position displacement of CDM particles corresponds to a Gaussian random realization of the CDM power spectrum $P_c(k, z_i)$; 2.) the initial position displacement of HDM particles corresponds to a Gaussian random realization of the HDM power spectrum $P_h(k, z_i)$; 3.) the initial velocities of both CDM and HDM particles, contributed by gravitational clustering, are assigned by the usual Zel'dovich approximation, using a total density power spectrum $P(k, z_i) = [\Omega_{\rm HDM}\sqrt{P_h(k,z_i)} + \Omega_{\rm CDM}\sqrt{P_c(k,z_i)}]^2$ and a flat universe dominated by CDM; 4.) in realizing the above three steps, the random phases of both CDM and HDM perturbations are kept the same, so that they correspond to a single random process; and 5.) each HDM particle is given a thermal motion randomly orientated and drawn from the Fermi-Dirac distribution [Eq.(1)]. We believe that all of the above requirements are correct except the third one. The third is not generally true because of the free-streaming motion of neutrinos. However, because the Jeans wavenumber of neutrinos is $1.9\,h\,{\rm Mpc}^{-1}$ at $z_i = 8$, greater than the Nyquist wavenumber $k_N = 1.6\,h\,{\rm Mpc}^{-1}$ of the simulation, the free streaming motion should not be very important in our simulation and the method of generating the initial conditions should be valid. The test performed by Jing et al. (1993b) also supports that the method is a good approximation for simulations we do here.

As well known, the cluster formation is sensitive to the linear density perturbations on scales of $\sim 10\,h^{-1}{\rm Mpc}$. Obviously, our initial conditions can correctly describe the linear density perturbation on a wide range around the cluster scale. Furthermore, the high force resolution of the P$^3$M code enables us to follow the density evolution in the non-linear regime and unlike the previous PM simulations of the model, to accurately simulate internal structures of clusters around their cores. Therefore we believe the simulation is very suitable for studying the properties of clusters.

The simulation method (including simulation parameters) for the SCDM and LCDM model is exactly the same as that for the CHDM model except no HDM particles are involved. The model parameters are specified as stated previously. We have run five realizations for the LCDM model and three realizations for the SCDM model. Since there are about two times more rich clusters in the SCDM model than in the other two models at $z = 0$, the total number of simulated clusters in the SCDM model is about the same as other two models.

## 3. Identifying rich clusters

We identify clusters based on particle count $C_{cl}$ or equivalently mass $M_{cl}$ within a sphere of some specific radius $r_{cl}$. Choosing $r_{cl}$ is somewhat arbitrary, and in this work we take $r_{cl} = 1.5\,h^{-1}{\rm Mpc}$ in the comoving coordinate. Our identifying method is similar to White et al. (1987a) and JMBF93. First we use the *friends-of-friends* algorithm to find groups of CDM particles with the link parameter $b$ equal to 0.2 times the mean separation of CDM particles. Around the barycenter of each group, we draw a sphere of radius $r_{cl}$ and measure total mass $M_{cl}$ of CDM and HDM particles inside the sphere. These spheres are the candidates of our clusters. When the center separation of two spheres is less than $2r_{cl}$ (i.e., the two clusters overlap), the cluster with smaller $M_{cl}$ is deleted from the candidate list. Because the groups identified by the friends-of-friends algorithm usually have very irregular shapes (e.g., filament-like structures), it is possible that some rich clusters in dense supercluster regions (e.g., where two rich clusters belong to the same group) would be missed by the above procedure. Therefore we iterate the above steps for particles which have not been listed as cluster members until no further cluster candidate is found [As we will see, this step could produce false clusters which are actually the outskirts of rich clusters]. Finally only the clusters with mass $M_{cl}$ above a certain value $M_{lim}$ are kept in our cluster catalogue. $M_{lim} = 8.9 \times 10^{13} M_\odot$ for the models of the CHDM and the SCDM, and $M_{lim} = 1.7 \times 10^{13} M_\odot$ for the LCDM. The mass limits correspond to 20 CDM particles within a sphere of $r_{cl}$.

We have checked particle distributions around the clusters identified above. Although most of the clusters are regular with dense central density of particles, some of them, however, have very complicated and irregular shapes. A few of examples from one realization of the LCDM model are given in Figure 1 (these examples are typical in every model). The figure shows the particle distributions around cluster centers within radii $r_{cl}$ (dotted circles) and $2r_{cl}$ (solid circles in Fig. 1a and 1b) or $4r_{cl}$ (solid circles in Fig. 1c). The cluster in Fig. 1a consists of two massive halos separated by $\sim 3\,h^{-1}$Mpc. Because of a thin bridge (left of the cluster center) between the two halos, the halos are identified as one group by the *friends-of-friends* algorithm and its mass center lies between the two halos. Fig. 1b shows a cluster whose center deviates from the densest region (dashed circle). If the cluster center moved to the densest spot, the mass within the radius $r_{cl}$ would be less than the mass within the dotted circle. However, the cluster finding algorithm is designed to find out the most massive regions of *radius $r_{cl}$*. Fig. 1c shows a cluster which actually is the outskirt of a much richer clusters. This false cluster is produced by the iterated searching for clusters from non-cluster members. These examples reflect the complicated distribution of particles around clusters. Such irregular clusters can hardly be avoided by simply adjusting the parameters $b$ and $r_{cl}$ in the cluster-finding algorithm.

These complicated clusters bring about difficulties in our measuring their velocity dispersions. As a first thought, we would measure velocity dispersion of particles around cluster centers. This will certainly give misleading results.

Massive halos must be located at gravitational potential minimums. The positions of the minimums generally are the densest spots of particle distribution. If we have positions of these minimums, then we can easily calculate the velocity dispersion around the minimums (the massive halos). From our tests, it seems much more robust to find potential minimums than to find density maximums.

Our identification of the potential minimums has been done around the clusters identified previously. Around each cluster, we place a cube of side length $12.8\,h^{-1}$Mpc with the cluster center overlapping the cube center. In the cube, a grid of $64^3$ uniform meshes is placed. Then we accumulate mass density values for the meshes from particles within the cube. We use the Gaussian kernel

$$W(r,s) = \frac{1}{(2\pi)^{3/2} s^3} \exp \frac{-r^2}{2s^2}, \qquad (2)$$

to get the density values, where $s$ is the smoothing length. To have the density field reasonably resolved and smoothed, we use a spatially varying smoothing length for each particle. The smoothing length $s_i$ of particle $i$ is taken to be its local mean separation $d_i$ of particles, $s_i = d_i$. The $d_i$ of a CDM particle is calculated by counting its 5 nearest CDM particles. Similarly we calculate $d_i$ for HDM particles by counting their HDM neighbours. Because we are collecting density values on the grid, we require $s_i$ not be less than the cell size, i.e. the density field is smoothed at least on one cell size. Then the density value on arbitrary cell $j$ is given by

$$\rho_j = \sum_i m_i W(r_{ji}, s_i), \qquad (3)$$

where $r_{ij}$ is the separation between cell $j$ and particle $i$ of mass $m_i$. The summation in Eq.(3) is over all particles with $r_{ij} < 5s_i$. We have also tried other choices of the smoothing lengths $s_i = 0.7 d_i$ and $s_i = 1.4 d_i$, and we find that our results are quite robust to these changes.

Gravitational potential on the grid is then obtained, like in the N-body simulation, by the FFT techniques. To eliminate the boundary effects around the grid, we consider only the central cubic volume of side length $7.7\,h^{-1}$Mpc (which is about $5 r_{cl}$). One cell is identified as a potential minimum if its potential value is smaller than those of its all 26 neighbors. Because the clusters are separated by not less than $2 r_{cl}$, some potential minimums can be identified for more than one time and we have corrected for this.

Success of the above procedure is illustrated in Fig. 1. For the cluster of Fig. 1a, we have identified two potential minimums which are labeled by dashed circles. For the cluster of Fig. 1b, we have identified one minimum (again dashed circle) within the radius $r_{cl}$ [A small halo at lower-right side of the dotted circle is also identified as a minimum]. For the *central* cluster of Fig. 1c, we have not found any minimum, i.e. there is only one minimum within the solid circle. All this shows that the procedure works quite successfully.

## 4. Abundance of clusters

In this section, we will calculate abundance of clusters in the simulations and compare the predictions with the observations. As stated in §1, there are three relevant observations: the mass function, the velocity dispersion function and the temperature function.

### 4.1 The mass function

BC92 presented a local cumulative mass function, $n(>M)$, which is defined as the number of objects per unit volume with mass larger than $M$. They measured mass $M$ within a sphere of radius $1.5\,h^{-1}$Mpc using some empirical relations (e.g., relationships of optical luminosity and mass, velocity dispersion and mass, etc.). Because of this definition of mass, we will use cluster samples from the *friends-of-friends* algorithm, since these samples collect clusters based on mass within radius $1.5\,h^{-1}$Mpc. It seems not easy to measure the function at moderate redshift (e.g., 0.5) in near future, we present model prediction only at the present time ($z = 0$). Figure 2 shows the mass function (MF) of the CHDM model, compared with the statistical data of BC92. [Uncertainty in the N-body MFs (and VDFs below) can be judged by the number of clusters involved which is the product of $n(M)$ ($n(>\sigma_v)$) and the total simulation volumes.] The model MF has the same shape as the observed MF, and the two MFs are consistent within $\sim 1\sigma$ error. The model MF is roughly at the $1\sigma$ upper limit of the observed MF. Biviano et al. (1993) and Giuricin et al. (1993) argued for a higher MF based on their analyses of cluster velocity dispersion data. This may make the CHDM model fit the observation better. Compared with JMBF93, the present MF is lower so is in better agreement with the observation. The main reason for the discrepancy is likely that different linear power spectra are used: JMBF93 used the spectrum of Holtzman (1989) and this study uses the spectrum of KHPR93. When the two spectra are normalized by the COBE quadropole, the spectrum of Holtzman is higher by $\sim 20\%$ on rich cluster scale. Since the spectrum of KHPR93 is more accurate, the result of this paper is more accurate. We noticed that quite recently Klypin and Rhee (1993) has calculated MF for this model, and our MF is in good accordance with their MF. For comparison, the LCDM model predicts $\sim 50\%$ less clusters than the CHDM model but agrees better with the BC92 data. The SCDM predicts more clusters than the CHDM model and the BC92 observation. Our results of the LCDM and SCDM models agree well with BC92.

### 4.2 The velocity dispersion function

Based on their survey of $R \geq 1$ Abell clusters within $z < 0.05$ and the CfA redshift survey of galaxies, ZGHR93 worked out a cumulative velocity dispersion function (VDF) $n(>\sigma_v)$ of groups and clusters. VDF is defined in a way similar to MF. Since the velocity dispersion $\sigma_v$ can be observed directly (on the contrary, mass measurement is usually model-dependent), VDF provides a good test of theoretical models. Observed $\sigma_v$ comes from the measurement of galaxy redshift. We assume here that there is no velocity bias between galaxies and underlying dark matter, so that the model $\sigma_v$ can be calculated from the peculiar velocity of dark matter particles.

Around the potential minimums of all simulated clusters, we calculated velocity dispersion $\sigma_v$ from the peculiar velocity of CDM particles within a sphere of physical radius $r_v = 0.5\,h^{-1}$Mpc. Choosing the $r_v$ is some arbitrary, but our results do not change if we take $r_v = 0.75\,h^{-1}$Mpc. To calculate the VDF, we need a simulated sample of clusters *complete* to certain velocity dispersion. Although there is strong correlation between mass $M_{cl}$ (or count $C_{cl}$ of CDM particles) and $\sigma_v$, dispersion of the $M_{cl}$-$\sigma_v$ or the $C_{cl}$-$\sigma_v$ relation is quite large. In Figure 3, we show this relation of our simulations. Since one cluster and its neighboring region could have more than one potential minimum (see Fig. 1a) or equivalently more than one $\sigma_v$, in Fig. 3 we plot only the highest value of $\sigma_v$ (i.e. main halo) for each cluster. Although there is large spread in the $C_{cl}$-$\sigma_v$, the upper envelope of the relation provides reliable criterion to define a complete sample. If we search for potential minimums around all clusters of count greater than $C_{lim}$, the sample of potential minimums must be complete for $\sigma_v > \sigma_{lim}$. We use this criterion (i.e. the upper envelope of the $C_{cl}$-$\sigma_v$) to obtain our complete samples of velocity dispersion. For the CHDM and the LCDM, $\sigma_{lim} = 700\,\mathrm{kms}^{-1}$, and for the SCDM, $\sigma_{lim} = 866\,\mathrm{kms}^{-1}$.

As shown in our future paper (Jing et al. 1994), some clusters in these hierarchical models exhibit substructures, as in the real observations (Forman et al. 1981; Jones & Forman 1992 and references therein). These clusters have more than one potential minimum in their internal regions, say within radius $1\,h^{-1}$Mpc. In real observations, such a cluster is most likely regarded as an individual cluster. To mimic the observations, if there are several minimums separated by less than $d = 1\,h^{-1}$Mpc, we only remain the minimum of the highest velocity dispersion. Since only a small fraction of clusters (less than 10%) show bimodal substructures (Jing et al. 1994) and $n(\sigma_v)$ is a strong decreasing function of $\sigma_v$, our results of $n(\sigma_v)$ are essentially insensitive to how we account for the substructures. If we use $d = 2\,h^{-1}$Mpc or $d = 0.5\,h^{-1}$Mpc, we get essentially the same results. However, by selecting the potential minimums, we can properly measure velocity dispersions for clusters with substructures.

Figure 4a presents our VDF determination for the CHDM model. We measured the VDF at four different redshifts as shown in the figure. As shown in §4.3, the VDF at $z \approx 0.5$ could be an important test for the model. Here we first compare the model prediction with the CfA data of ZGHR93 at $z \approx 0$. ZGHR93 analyzed the line-of-sight velocity dispersions, so we have shifted their data by a factor of $\sqrt(3)$ in the figure to comapre with the model predictions of 3-D velocity dispersions. The figure shows that the model can nicely fit the observation for $\sigma_v \gtrsim 700\,\mathrm{kms}^{-1}$. We also noticed a tendency that the model predicts more clusters of low velocity dispersion ($\leq 500\,\mathrm{kms}^{-1}$) than the observation. However, the observation at $\sigma_v \leq 500\,\mathrm{kms}^{-1}$ could be seriously underestimated for two reasons. First, as ZGHR93 cautioned, their calculations of group velocity dispersions $\lesssim 500\,\mathrm{kms}^{-1}$ are often underestimates. Second, they measure $\sigma_v$ only for CfA groups with $\geq 5$ group members. For $\sigma_v \lesssim 500\,\mathrm{kms}^{-1}$, a quite fraction of groups could have members less than 5.

For comparison, we show the VDF predictions of the LCDM model and the SCDM model in Figures 4b and 4c. The VDF of the LCDM at $z = 0$ is consistent with the CfA observation within $1\sigma$ error for $\sigma_v \lesssim 1500\,\mathrm{kms}^{-1}$. For larger $\sigma_v$, the model abundance is not well determined since there are only a total of 8 clusters of $\sigma_v > 1500\,\mathrm{kms}^{-1}$ in the simulation. The general trend is that the model prediction is at the $1\sigma$ lower limit of the CfA observation for $\sigma_v > 1000\,\mathrm{kms}^{-1}$. A LCDM model with a smaller $\Omega h$ than that we use could fit the observation better. The SCDM model predicts more clusters than the CfA observation except at $\sigma_v \approx 2000\,\mathrm{kms}^{-1}$ where the model prediction is in good agreement with the observation. The model is rejected at a very high confidence level ($> 6\sigma$) by the observational data $n(> 866\,\mathrm{kms}^{-1})$ and $n(> 1500\,\mathrm{kms}^{-1})$. Reducing the normalization of the power spectrum (i.e., reducing $\sigma_8$) may reduce formation of clusters thus resolving the problem of too many clusters. But since $n(>\sigma_v)$ is very steep at high $\sigma_v \approx 2000\,\mathrm{kms}^{-1}$, reducing $\sigma_8$ probably leads to a deficit at $\sigma_v \approx 2000\,\mathrm{kms}^{-1}$. And the model would conflict more severely with the COBE measurement of $Q$.

### 4.3 Cluster abundance at moderate redshift

The CHDM model and the LCDM model can equally well explain the abundance of clusters at redshift $z = 0$ (§4.1, §4.2 and §4.4). This is in accordance with previous studies which showed that the two models can equally well explain many other observations on large scale structures, including the two-point correlation functions of galaxies, the cluster-cluster correlation function, the pairwise velocity dispersion of galaxies on $\sim 1\,h^{-1}\mathrm{Mpc}$, the large scale bulk motions as well as the COBE measurement of the MBR quadropole. Indeed both models are competitive and are attracting more and more attention. It is very important to find tests to discriminate between the two models.

The abundance of clusters at moderate redshift $z \approx 0.5$ can become such a discriminator. As shown in Fig. 4, the abundance of clusters decreases very rapidly with redshift $z$ between $z = 0.5$ and $z = 0$ in the CHDM model, but evolves little in the LCDM model. The result is as expected because the linear perturbation continues growing in the CHDM model so that new collapsing objects (new clusters) are formed. In contrast, in the LCDM model the linear perturbation growth almost stops after $z = 0.5$, so few new clusters form between $z = 0.5$ and $z = 0$. In the CHDM model, the abundance of clusters at $z = 0.5$ is about one fourth of the present abundance for $\sigma_v \gtrsim 700\,\mathrm{kms^{-1}}$ and about one tenth for $\sigma_v > 1200\,\mathrm{kms^{-1}}$. Measurement of cluster abundance at $z \sim 0.5$ will provide a sensitive test to the models.

The CHDM model appears in severe conflict with the Dressler and Gunn measurement of velocity dispersions (Gunn 1990) for distant clusters in the GHO catalogue. In the catalogue, there are five clusters which appear as rich as the Coma cluster in about $3 \times 10^6\,h^{-3}\mathrm{Mpc}^3$ comoving volume at $z \approx 0.8$. Dressler and Gunn further measured $\sigma_v$ for two of these clusters, and found $\sigma_v \approx 1650\,\mathrm{kms^{-1}}$ for both of them, which confirms that the clusters have richness similar to the Coma. This means that the density of rich clusters with $\sigma_v \gtrsim 1600\,\mathrm{kms^{-1}}$ is as high as $1.6 \times 10^{-6}\,h^3\,\mathrm{Mpc}^{-3}$ at $z = 0.8$. In contrast, the CHDM model predicts a density of the $\sigma_v > 1500\,\mathrm{kms^{-1}}$ clusters $2 \times 10^{-7}\,h^3\,\mathrm{Mpc}^{-3}$ at $z = 0.5$. Our simulation has not output results at $z > 0.5$, but the cluster abundance at $z = 0.8$ is certainly lower than that at $z = 0.5$. Using the P-S formula as described in §5, we find that $n(> 1500\,\mathrm{kms^{-1}})$ at $z = 0.8$ is about one tenth of that at $z = 0.5$. As the result, the abundance of clusters predicted by the CHDM model is lower by 80 times than the observation. Based on the Poisson statistics, probabilities of the CHDM model having 2, 3, 4 and 5 clusters of $\sigma_v > 1500\,\mathrm{kms^{-1}}$ in a comoving volume $3 \times 10^6\,h^{-3}\mathrm{Mpc}^3$ at $z = 0.8$ are $1.7 \times 10^{-3}, 3.4 \times 10^{-5}, 5.1 \times 10^{-7}$ and $6.1 \times 10^{-9}$ respectively. Taking these face values, the CHDM model can be ruled out at a very high confidence level.

### 4.4 Temperature function of clusters

Another related statistic is the temperature function $n(T)$ (TF), which is defined as the mean number of clusters expected in a unit volume and in a unit gas temperature range. Because our simulations are not hydrodynamical, we assume that gas temperature $T$ is equal to the virial temperature $T_{vir}$ of the cold dark matter in dense cluster cores, i.e.,

$$\frac{3}{2}kT = \frac{1}{2}\mu m_p \sigma_v^2, \qquad (4)$$

where $\mu$ the mean molecular weight in amui and $m_p$ is the proton mass. While the assumption is based on the fact that the gas and the dark matter are in the same potential well, more unambiguous and direct support for this assumption comes from the recent hydrodynamical simulations of rich clusters (Thomas & Couchman 1992; Evrard 1990), which showed that the difference between the two temperatures is less than 10% in the cluster cores. According to the observations of Edge et al. (1990), we take $\mu = 0.58$ for intracluster gas.

Edge et al. (1990) and Henry & Arnaud (1991) have constructed the TF for local

clusters ($z = 0$) based on data from the *Einstein* and *EXOSAT* x-ray satellites. Although the samples they used are similar, their results differ by a factor 3 in $n(T)$. Using the P-S formula and assuming a relation between cluster virial mass and cluster gas temperature, Bartlett & Silk (1993) calculated the TFs for several theoretical models, including the CHDM and the LCDM models. Klypin & Rhee (1993) analyzed the local TF for the CHDM model based on their PM simulation and on the assumption $T \propto \sigma_v^2$.

Figure 5a presents our TF calculation for the CHDM model. We analysed the TF not only for $z = 0$, but also for three other earlier times. Our TF is calculated in this way. First we order all clusters of each model from high $T$ to low $T$. We bin the clusters so that in the first bin (highest $T$) there are 10 clusters, in the second bin there are 20 clusters, ..., and so on, until there are no enough clusters in the last bin according to this rule. Then the TF is estimated in these bins. Our TF at $z = 0$ is in good agreement with the observational result of Henry & Arnaud (1991), but higher by a factor $\sim 3$ than that of Edge et al. (1990) at high temperature $T \sim 8 \text{KeV}$. Compared with the previous theoretical calculations, our local $n(T)$ agrees quite well with Klypin & Rhee (1993), but is a bit lower (about a factor 2) than that of Bartlett & Silk (1993). The main cause for the theoretical discrepancy might be due to different transfer functions used in the different works (see §1 for a discussion).

For comparison, the SCDM model predicts too many clusters both at low and at high temperatures (Figure 5c). The LCDM model (Figure 5b) however predicts right amount of clusters compared with the data of Henry & Arnaud (1991). These results are consistent with our conclusions made in §4.1 and §4.2.

The CHDM model shows a very rapid evolution of the TF: the $n(T)$ at $z = 0.5$ is about one fifth of that at $z = 0$. In contrast, the LCDM model has very little evolution in $n(T)$ since $z = 0.5$. This is expected as explained in §4.3, since the gas temperature $T$ and the velocity dispersion $\sigma_v$ are simply related by Eq. (4). These quantitative predictions will be tested by some ongoing deep x-ray surveys of clusters.

Since cluster gas temperature $T$ and its x-ray luminosity $L_x$ are strongly correlated (e.g., Edge et al. 1990; Henry & Arnaud 1991; David et al. 1993), it appears that the rapid evolution of $n(T)$ of the CHDM is very compatible with the observed evolution of the luminosity functions found by Edge et al. (1990) and Henry et al. (1992). While the $T$-$L_x$ relation is governed by a lot of unknown physical processes (e.g., star formation in clusters, origin of intracluster medium, and formation of cluster cores) and its evolution with redshift is even unclear, we can not actually assess if the CHDM model and the LCDM model are consistent with the observed evolution of the cluster luminosity function. If $L_x(T, z)$ is a decreasing function of $z$, the LCDM model may be able to explain the evolution of the luminosity function.

At the present, there exist no observational data for determining the TF of x-ray clusters at $z \approx 0.5$. But it is still very instructive to compare our model prediction, based on the empirical $T$-$L_x$ relation, with deep X-ray survey of clusters without temperature determination. The EMSS can serve for this purpose. In the EMSS, there are six clusters with redshift $0.4 < z < 0.6$. Using the $V_{max}$ method as Henry et al (1992) determined x-ray luminosity functions for this survey, we find from these six clusters that the mean density of clusters with x-ray luminosity $L_{0.3-3.5}$ in the $0.3 - 3.5$ KeV band greater $4 \times 10^{44} \mathrm{ergs.s}^{-1}$ ($h = 0.5$, $q_0 = 0.5$) is about $1.2 \times 10^{-7} \, h^3 \, \mathrm{Mpc}^{-3}$. Assuming that the local empirical relation $L_{0.3-3.5} = 0.085(kT/\mathrm{KeV})^{2.4} \, 10^{44} \mathrm{ergs.s}^{-1}$ (Henry et al. 1992) is applicable to $z = 0.5$, the above estimation implies a density $1.2 \times 10^{-7} \, h^3 \, \mathrm{Mpc}^{-3}$ for clusters of $kT > 4.9 \mathrm{KeV}$. In our CHDM simulation of a total $10^7 \, h^{-3} \mathrm{Mpc}^3$ volume, we have two clusters with $kT > 4.9 \mathrm{KeV}$ at $z = 0.5$. So the CHDM is quite consistent with the EMSS observation.

The cluster density at $z = 0.5$ from the EMSS appears in contradiction with that of the GHO survey, unless the local $L_{0.3-3.5}$-$T$ relation is not valid at $z = 0.5$ and $L_{0.3-3.5}(T, z)$ is a *decreasing* function of $z$ (i.e., for clusters with a given temperature $T$, x-ray is dimmer at $z = 0.5$ than at present), or the temperature is not simply related to the velocity dispersion as Eq.(4). Although it is not easy to model the evolution of the $L_{0.3-3.5}$-$T$ relation, all simple models proposed for x-ray cluster formation (e.g., Kaiser 1986; 1989; Evrard & Henry 1991) predict an *increasing* function of $L_{0.3-3.5}(kT, z)$ with $z$ in an Einstein-de-Sitter universe. As stated earlier, hydrodynamical simulations have shown that Eq. (4) is a good approximation for rich clusters. Despite these simple model indications, to resolve the apparent discrepancy properly, it is essential to have the gas temperature and/or the velocity dispersion measured for the EMSS clusters at $z \approx 0.5$. The Canadian Network for Observational Cosmology (CNOC) is making effort to obtain velocity dispersion for these clusters (Carlberg et al. 1993). Detailed velocity dispersion measurement for those distant rich clusters in the GHO catalogue is also necessary to avoid any possible projection effects on $\sigma_v$ (e.g., Frenk et al. 1990; Bird & Beers 1993). Whether the CHDM model is consistent with the cluster abundance at $z = 0.5 - 1$ can be definitely answered when these observations are available.

## 5. The Press-Schechter formula as a function of velocity dispersion

The theory of Press & Schechter (1974) has been widely used to calculate the abundance of collapsed objects. In an Einstein-de Sitter universe of Gaussian primordial fluctuation, the theory predicts that the mean density of collapsed objects of virial mass $M$

at redshift $z$ is

$$n(M,z)dM = -\sqrt{\frac{2}{\pi}}\frac{\rho_0}{M^2}\frac{d\ln\sigma(M)}{d\ln M}\frac{1.68(1+z)}{\sigma(M)}\exp(-\frac{[1.68(1+z)]^2}{2\sigma^2(M)})dM \qquad (5)$$

where $\sigma(M)$ is the present linearly evolved *rms* density perturbation on scale $M$, and $\rho_0$ is the present mean density of the universe. Here 'collapsed objects' are defined as overdenisty regions of the mean density contrast 178 relative to the background. The mass and physical radius $r_{vir}$ of a collapsed object at $z$ can be related to its initial comoving radius $r_0$ (in present unit):

$$M = \frac{4\pi}{3}\rho_0 r_0^3; \quad r_{vir} = \frac{r_0}{178^{1/3}(1+z)} \qquad (6)$$

N-body simulations have shown that density profiles of the collapsed objects are approximately power laws $\rho(r) \approx r^{-\gamma}$, with $\gamma \approx 2$. If the objects are in equilibriums within spherical 'isothermal' potential wells, their velocity dispersions $\sigma_v$ can be related to $r_0$ by

$$\begin{aligned}\sigma_v^2 &= \frac{\gamma-\beta(\gamma-1)}{(\gamma-2\beta)\gamma}\frac{3GM}{r_{vir}} \\ &= \frac{[\gamma-\beta(\gamma-1)]178^{1/3}}{2(\gamma-2\beta)\gamma}3H_0^2 r_0^2(1+z)\end{aligned} \qquad (7a)$$

where $\sqrt{1+\beta}$ is the ratio of the tangential velocity dispersion to the radial velocity dispersion. The first factor on the right hand side of Eq.(7a) depends on parameters $\gamma$ and $\beta$, both of which are not well determined and may differ for different clusters and for different galaxy formation models. Let the uncertain factor be denoted as an unknown parameter $c_\sigma^2$, Eq.(7a) reads as

$$\sigma_v = c_\sigma\sqrt{3}H_0 r_0(1+z)^{1/2} \qquad (7b)$$

The abundance of collapsed objects as a function of $\sigma_v$ can then be expressed through Eq. (5):

$$n(\sigma_v,z)d\sigma_v = -\frac{3\times 1.68(1+z)}{(2\pi)^{3/2}r_0^3\sigma(r_0)}\frac{d\ln\sigma(r_0)}{d\ln r_0}\exp(-\frac{[1.68(1+z)]^2}{2\sigma^2(r_0)})\frac{d\sigma_v}{\sigma_v} \qquad (8)$$

The Press-Schechter formula as a function of mass $M$ [Eq.(5)] has been shown to be quite successful for a wide range of $M$ by N-body simulations of hierarchical clustering (Efstathiou et al. 1988; Bond et al. 1991). However it is much less clear whether Eq.(8) is similarly successful and if so, how much the parameter $c_\sigma$ is. As can be easily seen from the previous section, Eq. (8) can be very useful in linking theories to observations. NW88 first suggested $c_\sigma = 1.18$, which is equivalent to assuming $\gamma = 2$ and $\beta = 0$ in Eq.(7a). Comparing the prediction with the CDM simulation of White et al. (1987b), they find that Eq.(8) with the above $c_\sigma$ provides a reasonable fit to the simulation data for $\sigma_v \lesssim 500 \, \mathrm{kms^{-1}}$ (which their simulation can explore). Recently Carlberg et al. (1993) suggested another value $c_\sigma = 1.10$ based on their examination of largest collapsed halos in their N-body simulations (it is a pity that they did not give more details). Our carefully constructed samples of $\sigma_v$ can provide another ideal test to the theory. Because our simulations explore much larger halos ($\sigma_v \geq 866 \, \mathrm{kms^{-1}}$ in the SCDM model) than White et al. (1987b) did, our test is complementary to that of NW88. Furthermore the CHDM simulation can tell whether the parameter $c_\sigma$ depends on the shape of the power spectrum.

In Figure 6, we compare the P-S predictions of Eq. (8) with our simulation VDFs of the SCDM and the CHDM models. The comparison is made at two epochs $z = 0$ and $z = 0.5$. Because the neutrino free-streaming motion at $z \leq 8$ is not important for the formation of the halos of $\sigma_v \geq 700 \, \mathrm{kms^{-1}}$, the transfer function of the CDM at $z = 0$ is used for the P-S calculation of the CHDM model. We have tried three values, 1.1, 1.2 and 1.3 for the parameter $c_\sigma$. Generally a larger $c_\sigma$ gives a larger VDF. Although the P-S theory predicts a slightly steeper VDF than the N-body simulations, overall the P-S formula provides a good analytical approximation for both models. The simulation VDFs lie within the P-S predictions of $c_\sigma = 1.1$ and of $c_\sigma = 1.3$. Clearly for most calculations, $c_\sigma = 1.2$ is a preferred value.

## 6. Conclusions and further discussions

In this paper, we have presented the abundance of clusters at $z \leq 0.5$ for the CHDM model. The results are compared with the available observational data as well as with the models of the SCDM and the LCDM. Our results are:

1. The CHDM model is in good agreement with the local abundance of clusters ($z = 0$), including the mass function, the velocity dispersion function and the x-ray temperature function. When we compared the model predictions with the observed VDF and TF, we had assumed respectively that the velocity dispersion of galaxies is equal to the velocity dispersion of the CDM (i.e. no velocity bias) and that the gas temperature in clusters is equal to the virial temperature of the CDM [Eq.(4)]. The first assumption has been debated recently by a number of authors (e.g., Couchman & Carberg 1992), but there is no general consensus whether there is velocity bias and how much it is. The second assumption is justified by the hydro-N-body simulations (Thomas & Couchman 1992; Evrard 1990) which showed that Eq.(4) is valid for rich clusters within $\sim 10\%$ precision in $T$. For comparison, the LCDM model predicts a right amount of clusters at $z = 0$ as the CHDM, but the SCDM model predicts too many clusters of $\sigma_v \leq 1500 \, \text{kms}^{-1}$ compared with the observations.

2. The CHDM model shows a very rapid evolution of the abundance of clusters in the recent past: the abundances of clusters with $\sigma_v \gtrsim 700 \, \text{kms}^{-1}$ and with $\sigma_v \gtrsim 1200 \, \text{kms}^{-1}$ at $z = 0.5$ are only about 1/4 and 1/10 of the present abundances, respectively. On the contrary, the LCDM model shows little evolution of clusters since $z = 0.5$. Therefore the observation of cluster abundance at $z \approx 0.5$ can serve a very promising test to discriminate between these two interesting models both of which have been found in good agreement with the observational data of local galaxy clustering and of the microwave background radiation. The test is expected to be available in the near future with the launch of AXAF which will determine temperature function for clusters of $z < 0.5$. The ROSAT survey will determine reliably the x-ray luminosity function of $z < 0.5$ clusters which can be transformed to temperature function based on some empirical relations between the luminosity and the temperature.

3. Compared with the optical observation of distant clusters at $z = 0.8$ by Gunn and his collaborators, the CHDM model predicts a deficit of rich clusters ($\sigma_v \geq 1500 \, \text{kms}^{-1}$) which in mean density is about two magnitudes lower than the observation. However, another observation, the mean density of x-ray clusters of $L_{0.3-3.5} > 4 \times 10^{44} \text{ergs.s}^{-1}$ in the EMSS, seems in good accordance with the CHDM model if the local $L_x$-$T$ relation is valid at $z = 0.5$. We regard this result as tentative until detailed spectroscopic data (velocity dispersion and/or gas temperature) are available for the clusters in both samples.

4. We used our simulation data to test the P-S prediction of the halo abundance as a function of velocity dispersion $\sigma_v$. We find that the P-S expression [Eqs.(7-8)] is a reasonably good approximation if $c_\sigma$ is between 1.1 and 1.3.

## Acknowledgements

We are grateful to the referee, J.G. Bartlett, for comments. YPJ thanks the World Laboratory for a Scholarship. The work is partially supported by NSF contract INT 9301805.

# Figure captions

**Fig.1** Projected distributions of particles around centers of some clusters selected by the *friends-of-friends* algorithm in the LCDM model. Explanation is given in the text.

**Fig.2** The mass function of clusters in the CHDM model (solid line), compared with the statistical data of Bahcall & Cen (1992; circles) as well as with the predictions of the LCDM model (dotted line) and the SCDM model (dashed line).

**Fig.3** A scatter diagram of cluster counts of the CDM particles within radius 1.5 $h^{-1}$Mpc (determined by the *friends-of-friends* algorithm) and velocity dispersion $\sigma_v$ of its main halo. Potential minimums are searched for clusters with the counts larger than $C_{lim}$, thus the minimums (or the halos) searched are complete for $\sigma_v > \sigma_{\lim}$.

**Fig.4** The velocity dispersion functions of the three models at four different epochs since $z = 0.5$. Also plotted is the CfA data of Zabludoff et al. (1993) for local clusters and groups ($z \approx 0$).

**Fig.5** The temperature functions of the three models at four different epochs since $z = 0.5$. Also plotted are the local temperature functions ($z \approx 0$) observed by Edge et al. (1990) and by Henry & Arnaud (1991).

**Fig.6** The velocity dispersion functions from our simulations, compared with the Press-Schechter predictions with the parameter $c_\sigma = 1.1$, 1.2 and 1.3. Open and filled circles represent our simulation data at $z = 0$ and $z = 0.5$ respectively. The smooth lines are the P-S predictions of different $c_\sigma$. For each $c_\sigma$ choice, the upper line is for $z = 0$, and the lower one is for $z = 0.5$.

Figure 2

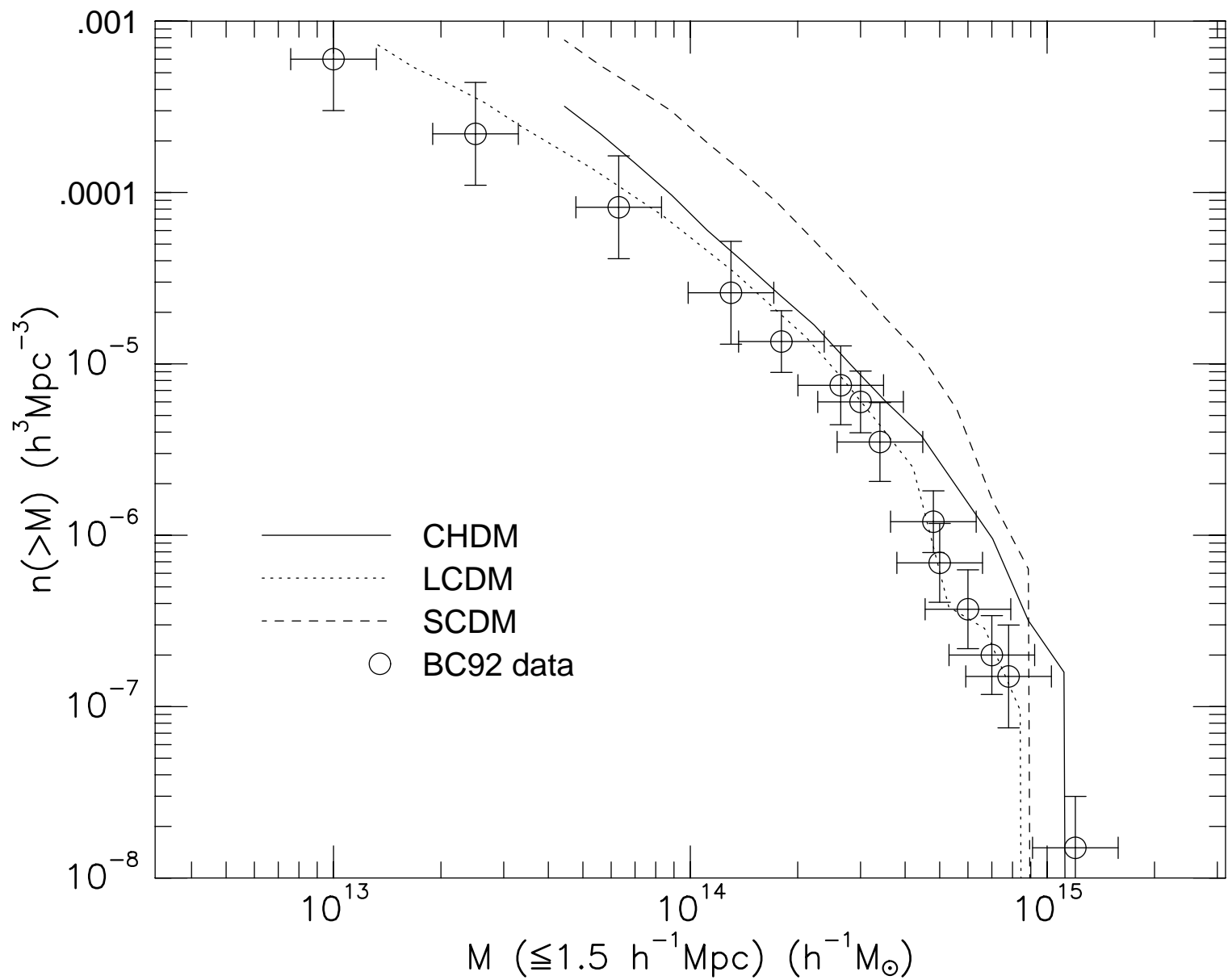

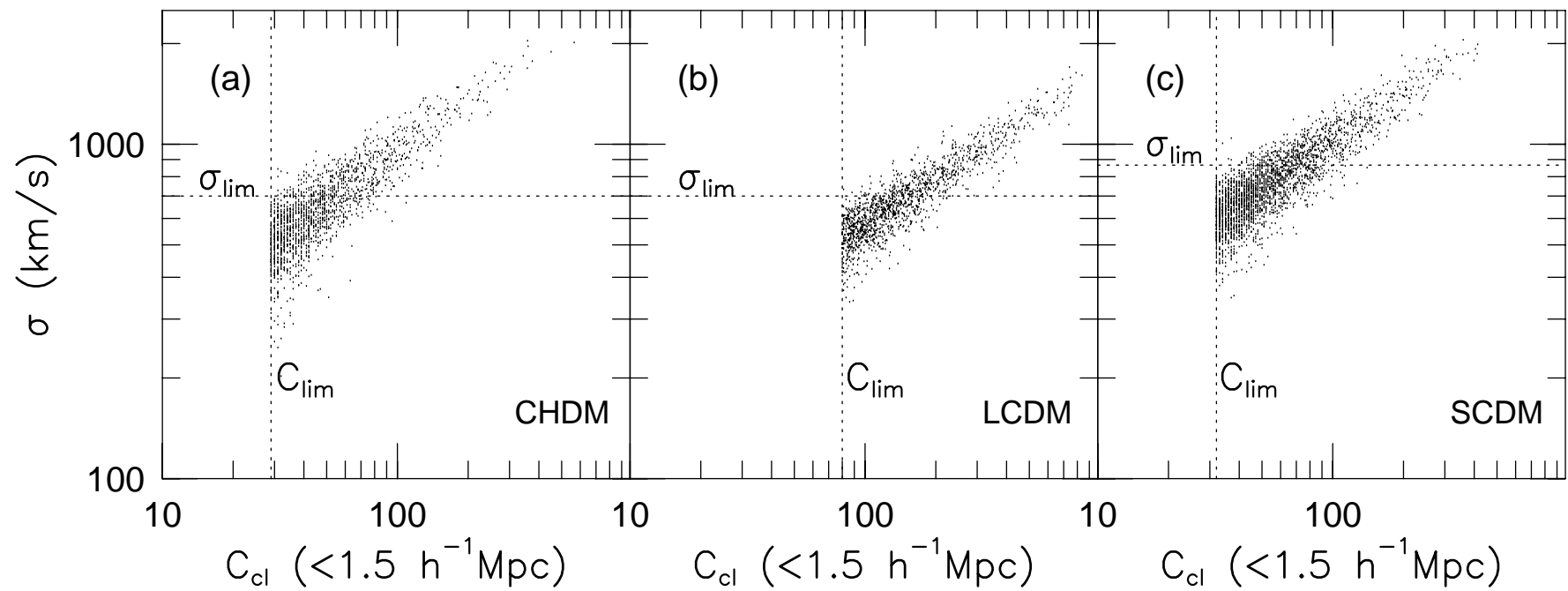

Figure 3

Figure 6

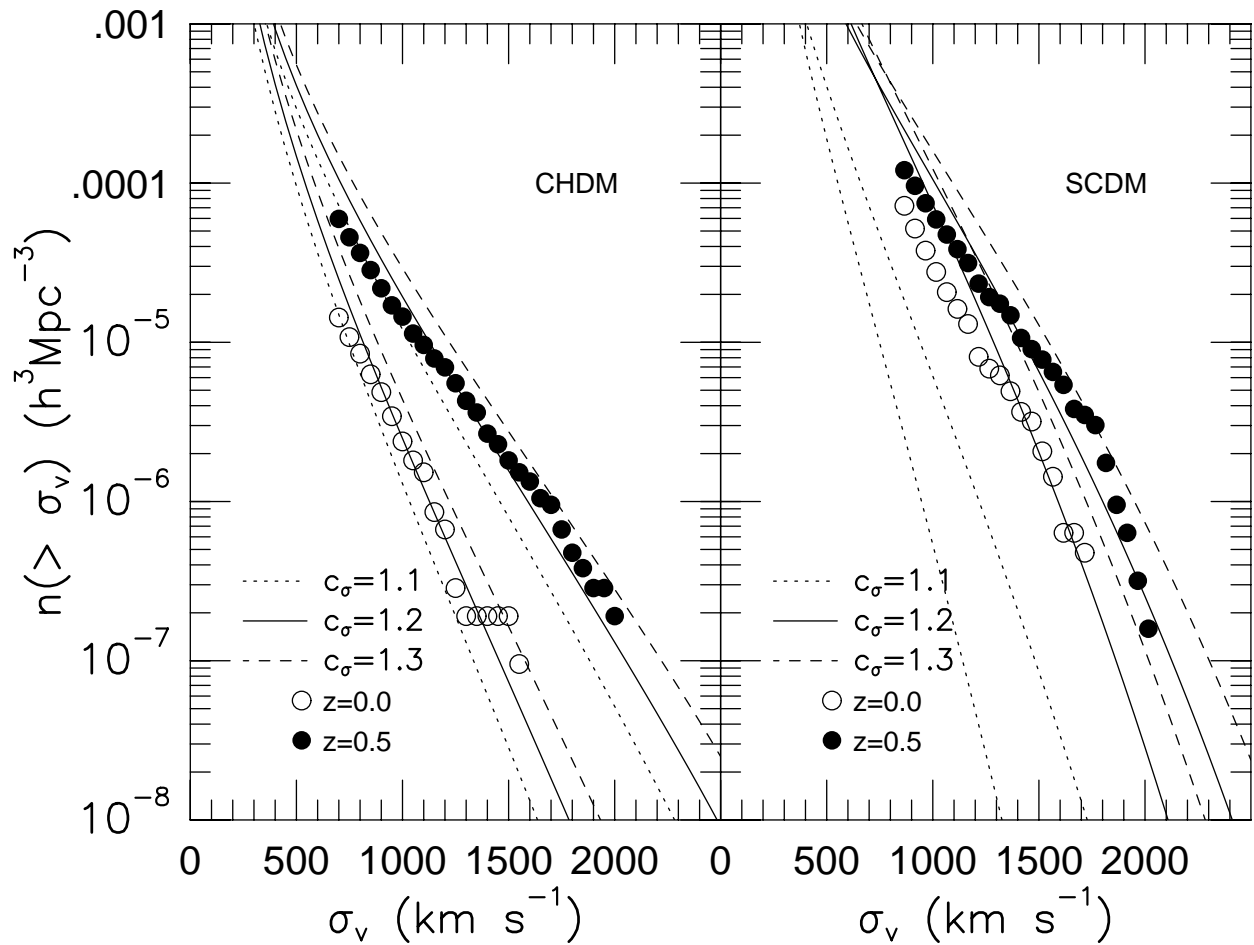

# Figure 5

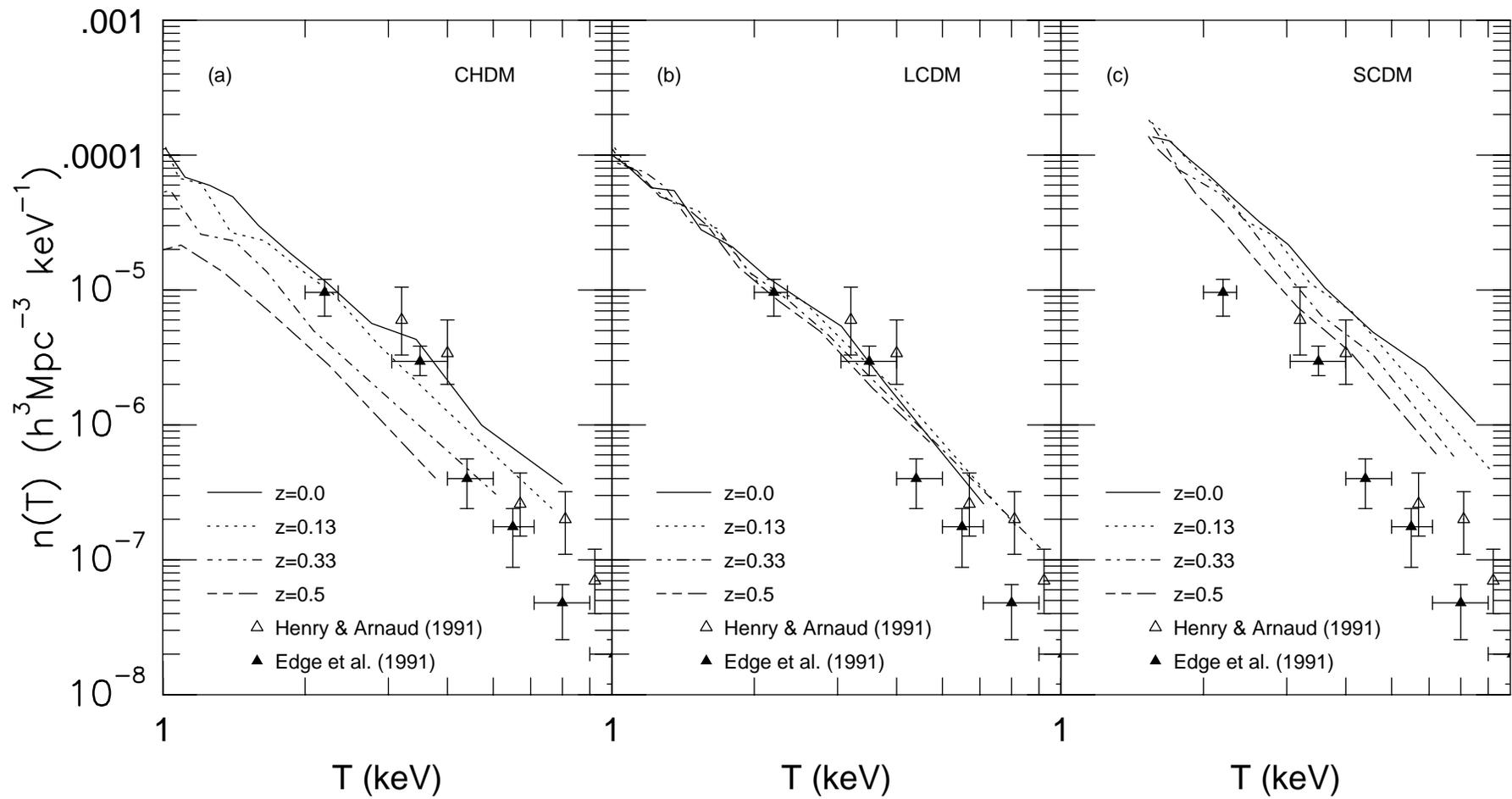

Figure 4

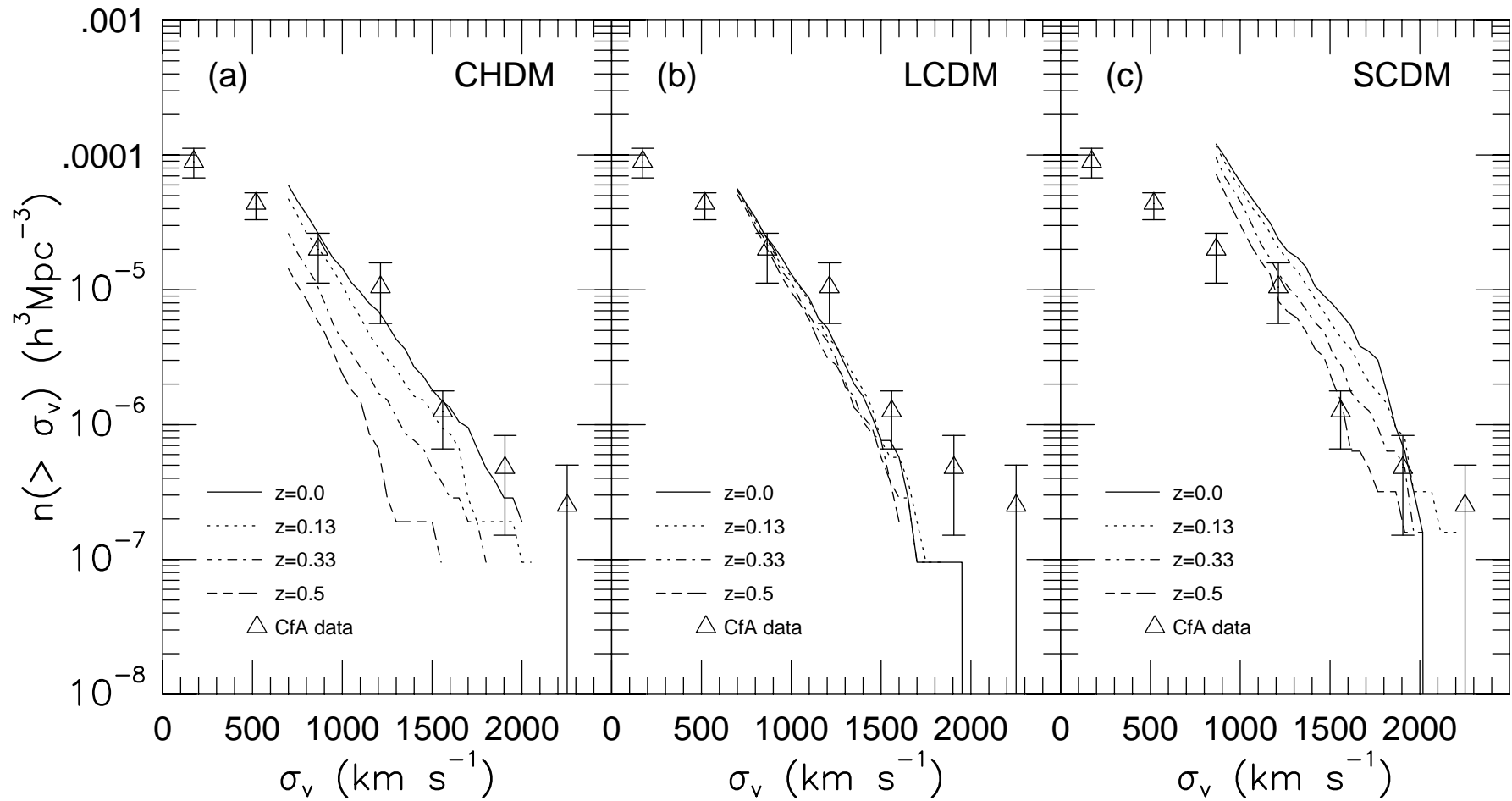